\documentclass[reprint,aps,superscriptaddress,amsmath,amssymb]{revtex4-2}

\usepackage{graphicx}
\usepackage{dcolumn}
\usepackage{bm}
\usepackage{color}
\usepackage[dvipsnames]{xcolor}
\usepackage{placeins}

\usepackage[normalem]{ulem}

\usepackage[linkcolor=blue,urlcolor=blue,citecolor=blue,colorlinks=true]{hyperref}

\newcommand{\rV}  {{\mathbf r}_\parallel}

\def\ESB{E_\textsc{sb}}
\def\GSB{g_\textsc{sb}}
\def\DTW{\Delta\tau}
\def\TFM{\tau_\textsc{fm}}
\def\TSI{\tau_\textsc{si}}
\def\TTT{\tau_\textsc{t}}
\def\TDW{\tilde\tau}
\def\TDD{\tilde\tau_{\textsc{3d}}}

\def\PSL{\Psi_\textsc{l}}
\def\DTW{\Delta\tau}
\def\ZL{z_{\rm L}}
\def\ZR{z_{\rm R}}
\def\XT{\xi^{t}}
\def\XR{\xi^{r}}
\def\Gv{\mathbf{g}}
\def\KKO{k_0}
\def\VVO{v_0}
\def\KKG{k_{\Gv}}
\def\VVG{v_{\Gv}}
\def\KDO{\dot k_0}
\def\KDG{\dot k_{\Gv}}
\def\GFA{G^{\textsc{1d}}_0}
\def\GFB{G^{\textsc{2d}}_0}
\def\Gv{\mathbf{g}}
\def\RRG{r_{\Gv}}
\def\RDG{\dot r_{\Gv}}
\def\TTG{t_{\Gv}}
\def\TDG{\dot t_{\Gv}}
\def\RRO{r_0}

\begin{document}
\title{Negative transit time in non-tunneling electron transmission through graphene multilayers}
\author{E. E. Krasovskii}
\affiliation{Universidad del Pais Vasco/Euskal Herriko Unibertsitatea, 20080 Donostia/San Sebasti\'{a}n, Basque Country, Spain}
\affiliation{Donostia International Physics Center (DIPC), 20018 Donostia/San Sebasti\'{a}n, Basque Country, Spain}
\affiliation{IKERBASQUE, Basque Foundation for Science, 48013 Bilbao, Basque Country, Spain}
\author{R.O. Kuzian}
\affiliation{Donostia International Physics Center (DIPC), 20018 Donostia/San Sebasti\'{a}n, Basque Country, Spain}
\affiliation{Institute for Problems of Materials Science NASU, Krzhizhanovskogo 3, 03180 Kiev, Ukraine}

\begin{abstract}
  Attosecond dynamics of electron transmission through atomically-thin crystalline films
  is studied with an {\em ab initio} scattering theory. The temporal character of the electron
  propagation through graphene multilayers is traced to the band structure of bulk graphite:
  In the forbidden gaps the wave packet transit time $\TTT$ saturates with thickness and in the
  allowed bands $\TTT$ oscillates following transmission resonances. Hitherto unknown negative
  transit time due to in-plane scattering is discovered in monolayers of graphene, h-BN, and
  oxygen. Moreover, Wigner time delay is found to diverge at the scattering resonances caused
by the emergence of secondary diffracted beams. This offers a way to manipulate the propagation
timing of the wave packet without sacrificing the transmitted intensity. The spatial reshaping
  of the wave packet at the resonances may help elucidate details of the streaking by an
  inhomogeneous field at the surface.
\end{abstract}

\maketitle

The question of how long does it take for a quantum particle to transit from one point
to another arises in various physical contexts, including nanotransport, photonics, and
photoelectron spectroscopy. There, one encounters paradoxical phenomena, such as the
Hartman effect~\cite{Hartman1962}---a fundamental aspect of wave propagation in a totally
reflecting medium whereby the traversal time of a wave packet across a finite slab is
independent of the slab thickness for sufficiently thick slabs. The implied unlimited
velocities inspired a lively theoretical discussion
\cite{Hauge1989,Landauer1994,Winful2006,Dumont2020,Field2022}
and triggered much experimental effort in optics
\cite{Steinberg1993,Enders1993,Spielmann1994,Balcou1997,Carey2000,Longhi2001}
strong-field ionization~\cite{Landsman2014,Eckle2008,Sainadh2019}, and atomic
tunneling~\cite{Ramos2020}.
The superluminality paradox has called for a refined understanding of the notion of propagation
velocity and a rigorous definition of the transit time~\cite{Hauge1989,Landauer1994,Winful2006}.
However, over the many decades the discussion has been limited to tunneling under a barrier,
while temporal paradoxes in classically allowed transmission have not been addressed. 

The progress in attosecond photoelectron spectroscopy of crystals~\cite{Cavalieri2007,
Schultze10,Neppl2012,Neppl15,Okell:15,Siek2017,Ossiander2018,Riemensberger2019} has given new
prominence to the question of how fast a quantum particle traverses a microscopic distance.
The measured phase shifts of the streaking spectrograms~\cite{Pazourek2015} are commonly
interpreted in terms of the escape time of photoelectrons originating at a depth of the order
of the mean free path~\cite{Cavalieri2007}. While there have been attempts to understand the
results in terms of group velocity, this is not strictly justified and completely fails in a
band gap. 

\begin{figure*}[t] %%%%%%%%%%%%%%%%%%%%%%%%%%%%%%%%%%%%%%%%%%%%%%%%%%%%%%%%%%%%%% F1
\centering
\includegraphics[trim = 0mm 0mm 00mm 0mmm,width=0.96\textwidth,clip]{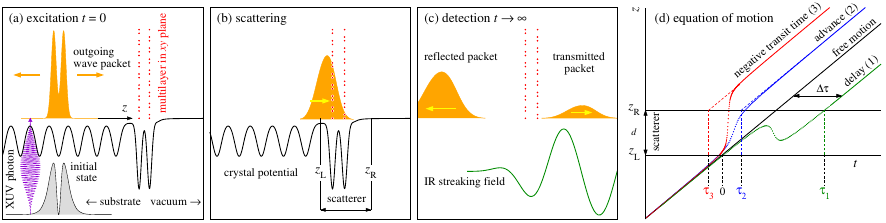}
\caption{(a)--(c)~Scattering configuration and a sketch of a relevant experiment.
    Excited (a), incident (b), reflected and transmitted (c) wave packets are shown by
    orange shading. (d)~Schematic equations of motion $z(t)$ of the incident and transmitted
  packets for positive (case 1) and negative (2 and 3) time delay $\DTW$.
  Straight solid lines show the asymptotic free motion, and dots symbolically indicate
  the region where $z(t)$ is undefined. The transit time is defined as the
  extrapolation of the $t\to\infty$ line to the right border of the scatterer, $z=\ZR$.
  Time delay~(1) is typical of transmission resonances, see Fig.~\ref{f:transit}.
  Time advance with a positive transit time~(2) is realized in the Hartman effect. Case~3
  is a new finding of the present work.
} 
\label{f:ein}
\end{figure*}

A rigorous way to proceed is to draw on the Wigner time delay $\DTW$, which is defined as the
difference in the time of arrival of a free particle and a scattered one in a region far from
the scatterer, see Fig.~\ref{f:ein}(d).
It can be obtained from the stationary states of the system as the energy derivative
of the scattering phase~\cite{Bohm1951,Wigner55,Smith60}. Here, we apply the phase-time
formalism to a tunneling-free propagation in experimentally accessible graphene multilayers.
$\DTW$ is calculated with a state-of-the-art band structure accuracy using the
augmented-plane-waves based scattering theory~\cite{Krasovskii2004}. We obtain realistic values
for the saturated transit time in the gaps and investigate the temporal characteristics of
transmission ($T$) resonances. Most important, we reveal the essential role of the lateral
scattering, which is inevitably present in realistic three-dimensional (3D) crystals, but
has been completely ignored in previous theoretical (exclusively 1D) studies.
Electron scattering from the monolayers of graphene and hexagonal boron nitride (h-BN) allows
to separate out the lateral scattering and reveals a striking effect: at the $T$-resonance
due to the emergence of the secondary diffraction beams the Wigner delay acquires
large negative values, which results in a negative transit time.

This effect can be measured in a laser streaking experiment: A subfemtosecond XUV pulse excites
a localized state just below the surface, Fig.~\ref{f:ein}(a), and the outgoing photoelectron
is scattered by the surface overlayer, Fig.~\ref{f:ein}(b), and further exposed in vacuum to
the laser streaking field synchronized with the XUV pulse, Fig.~\ref{f:ein}(c). The arrival
time is then inferred from the momentum transferred from the laser field to the electron
\cite{Cavalieri2007,Schultze10,Neppl2012,Neppl15,
Okell:15,Siek2017,Ossiander2018,Riemensberger2019}. Based on the time delay $\DTW$,
Hartman~\cite{Hartman1962}
introduced so-called transmission time $\TTT$, which implies that the distance $l$ traveled by
a wave packet can be divided into scattering region $d$ and free-motion region $l-d$, and that
the time to traverse the full length $l$ (arrival time~\cite{MUGA1995,ANASTOPOULOS2013}) equals
the sum of the two partial times. Although not rigorous~\cite{Hauge1997,Sokolovski2018},
this procedure is widely applied in analyzing the times of photoemission from surfaces or
strong-field ionization of atoms. Here, we ascribe by convention the width $d=na$ to an $n$ML
slab, where $a$ is the interlayer spacing. Then, $\TTT=\DTW+d/v_0$.

Let us consider an infinitely spectrally narrow wave packet normally incident on
a crystalline film located between $\ZL$ and $\ZR$ (outside this interval the electron
potential equals zero), see Fig.~\ref{f:ein}(b). The scattering wave function satisfies the
Schr\"odinger equation $\hat H\Psi=E\Psi$ with initial conditions implying the presence of
the incident and reflected wave(s) in the left half-space, $z<\ZL$, and transmitted wave(s)
in the right half-space, $z>\ZR$:

\begin{align}
  \Psi =
  \begin{cases}
e^{{ik_0z}}~+&\sum\limits_\Gv r_\Gv\, e^{i[-k_\Gv(z-\ZL)+\Gv\rV]},\; z\le\ZL,\\
           &\sum\limits_\Gv t_\Gv\, e^{i[+k_\Gv(z-\ZR)+\Gv\rV]},\; z\ge\ZR,
  \end{cases}
\label{eq:tr}
\end{align}
where $\rV$ is the surface parallel radius vector  and $\Gv$ are 2D reciprocal lattice
vectors. The incident wave vector is $k_0=\sqrt{2mE}/\hbar$, and $z$-projections of the
wave vectors of the secondary beams, $\Gv\ne0$, are $k_\Gv=\sqrt{2mE/\hbar^2-|\Gv|^2}$.
The scattering state $\Psi$ is calculated with the {\it ab initio} embedding method in
terms of augmented plane waves~\cite{Krasovskii2004}, 
see Appendix \ref{s:method}. % Sec.~I of Supplemental Material (SM)~\cite{Suppl}. 
Equation~(\ref{eq:tr}) introduces the transmission and reflection coefficients:
$t_\Gv = |t_\Gv|\exp(i\XT_\Gv)$ and $r_\Gv = |r_\Gv|\exp(i\XR_\Gv)$.
The scattering phase shift of the transmitted wave $\eta$ is related to the exit
phase $\XT_0$ as $\eta=\XT_0-dk_0$, where $d=\ZR-\ZL$.
The Wigner time delay $\DTW$ is the shift of the asymptotic $z\to\infty$ equation of motion
relative to that of the free particle, see Fig.~\ref{f:ein}(d),
and it equals the energy derivative of the phase
$\eta$~\cite{Bohm1951,Wigner55,Smith60}: $\DTW=\hbar d\eta/dE\equiv\dot\eta$.

If the slab has a symmetry operation $\hat s$ that swaps the planes $z=\ZL$ and
$z=\ZR$ then the following relation between reflection and transmission coefficients holds:

\begin{equation}
k_0 - \sum\limits_{|\Gv| < k_0} k_\Gv|t_\Gv + r_{\hat s\Gv}|^2 = 0,
\label{eq:IDCF}
\end{equation}
where the sum runs over the propagating beams $|\Gv| < k_0$, i.e., real $k_\Gv$.
The propagating secondary beams emerge at the energy $\ESB=\hbar^2\GSB^2/2m$,
where $\GSB$ is the magnitude of the shortest nonzero $\Gv$-vector. Below $\ESB$,
the current conservation law combined with Eq.~(\ref{eq:IDCF}) leads to the relation
$\XT_0-\XR_0=\pm\pi/2$, which is well known for 1D scattering~\cite{Falck1988}. For
several propagating beams it acquires a more complicated form of a cosine sum rule

\begin{equation}
\sum\limits_{|\Gv| < k_0} k_\Gv|t_\Gv r_{\hat s\Gv}|\cos(\XT_\Gv-\XR_{\hat s\Gv}) = 0.
\label{eq:IDCOS}
\end{equation}

An important identity relates the dwell time $\TDW$~\cite{Smith60}
\begin{equation}
\TDW= \frac{1}{v_0}\int\limits_{\ZL}^{\ZR}\!\!|\Psi(\rV,z)|^2\,dz,
\label{eq:DEFDWELL1D}
\end{equation}
to the Wigner time delay $\DTW$ in the 1D case~\cite{Smith60,Winful2003b}:

\begin{equation}
\TDW=\DTW+\frac{d}{v_0}+\frac{{\rm Im}\,r_0}{k_0v_0} =\DTW+\TFM-\TSI.
\label{eq:IDWINFUL}
\end{equation}
Here $v_0=\hbar k_0/m$ is the free-space velocity, so $\TFM=d/v_0$ is the free
particle transit time and $\TSI=\hbar\,{\rm Im}\,r_0/2E$ is a self-interference
delay~\cite{Winful2006}. In the 3D case the definition~(\ref{eq:DEFDWELL1D}) of
the dwell time must be modified to include the probability density contained in
the evanescent secondary beams in the whole space, see Eq.~(\ref{eq:Smith19}) %(S9)~\cite{Suppl}:
\begin{equation}
\TDD\equiv\TDW+\frac{1}{v_0}\sum\limits_{|\Gv|>\KKO}\frac{R_{\Gv}+T_{\Gv}}{2|k_\Gv|}, 
\label{eq:DEFDWELL3D}
\end{equation}
where $T_\Gv=|t_\Gv|^2$ and $R_\Gv=|r_\Gv|^2$ are the partial transmissivities and
reflectivities, respectively. The additional term is negligible well below $\ESB$,
and at $\ESB$ it diverges. Everywhere below $\ESB$, an analogue of Eq.~(\ref{eq:IDWINFUL})
holds $\TDD=\DTW+\TFM-\TSI$, and the general formula valid at all energies reads 
\begin{equation}\label{eq:GENID3D}
\TDD=\frac{1}{v_0}\sum\limits_{|\Gv|<\KKO}\frac{\hbar k_\Gv}{m}(T_\Gv\dot{\XT_\Gv}+
R_\Gv\dot{\XR_\Gv})+\frac{{\rm Im}\,r_0}{v_0k_0}.
\end{equation}
Equations~(\ref{eq:IDCF}) to~(\ref{eq:GENID3D}) are derived in Appendix \ref{app:ID}.
%Sec.~II of SM~\cite{Suppl}.

\begin{figure}[b] %%%%%%%%%%%%%%%%%%%%%%%%%%%%%%%%%%%%%%%%%%%%%%%%%%%%%%%%%% F2 
\centering
\includegraphics[trim = 0mm 0mm 00mm 0mmm,width=0.48\textwidth]{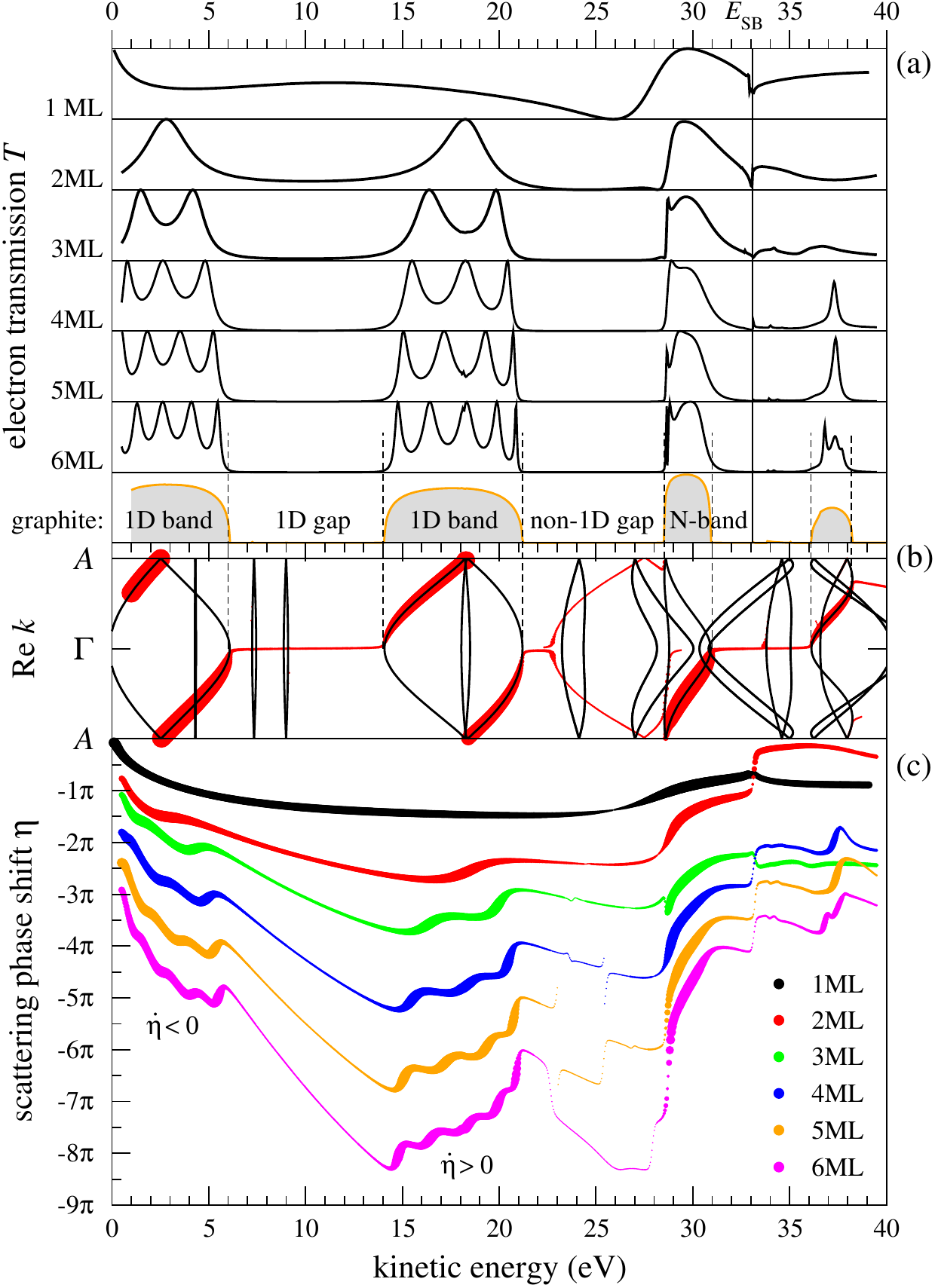} 
\caption{(a)~Transmission $T(E)$ through 1~to 6ML graphene and bulk graphite.
    $T$ ranges from~0 to~1 in all the panels. (b)~Black lines are the $A\Gamma A$
    band structure of graphite. Red circles are the conducting complex band structure
    \cite{Barrett05}: in the bulk, $\Psi$ is a sum of propagating and evanescent Bloch
    waves, and size of the circle is proportional to the current carried by the
    partial wave. Black lines not marked by red circles are irrelevant for transmission.
Red circles that do not mark any black lines are evanescent waves.
(c)~Scattering phase shift $\eta(E)$ for 1 to 6ML slabs. Size of the symbol is
proportional to the transmission amplitude $|t_0|$ of the main beam, see Eq.~(\ref{eq:tr}).
}
\label{f:CBS}
\end{figure}

\begin{figure}[h!] %%%%%%%%%%%%%%%%%%%%%%%%%%%%%%%%%%%%%%%%%%%%%%%%%%%%%%%%%% F5 NC3
\centering
\includegraphics[trim = 0mm 0mm 00mm 0mmm,width=0.48\textwidth]{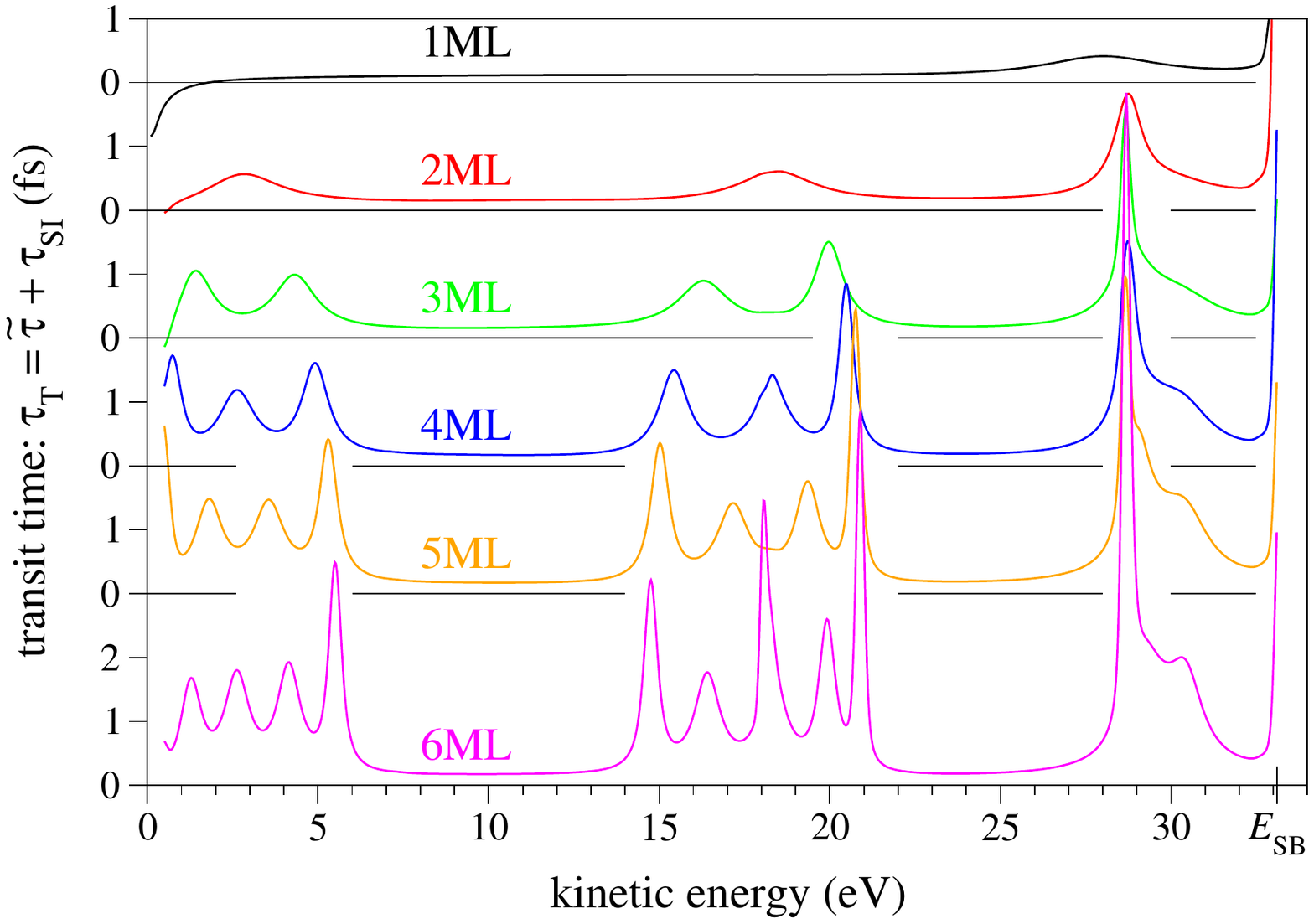} 
\caption{Transit time through 1~to 6ML graphene slabs. Spectra are shown up
to $\ESB$ because Eq.~(\ref{eq:IDWINFUL}) is used.}
\label{f:transit}
\end{figure}

Figure~\ref{f:CBS}(a) shows transmission spectra for 1 to 6ML graphene and
semi-infinite graphite. Clearly visible are conducting and reflecting intervals, evolving
respectively into bulk bands and forbidden gaps of graphite. Each of the two lowest bands hosts
$n-1$ spikes ($T=1$ resonances) for an $n$ML slab. They originate from interlayer
scattering~\cite{Hibino2008,Jobst-15,Jobst-16}, hence are referred to as 1D bands. At
the resonances, the transit time sharply peaks, see Fig.~\ref{f:transit}, similar to the
resonant tunneling through a double barrier~\cite{LIU1987}. At low energies the larger velocity
at the nuclei causes a faster propagation, as in the classical mechanics, so $\DTW$ is on
average negative, Fig.~\ref{f:CBS}(c). In the upper 1D band, the band structure effect
outweighs the classical acceleration leading to an overall positive slope of $\eta(E)$. 

The N-band around 30~eV is related to a scattering resonance due to coupling
of the in-plane and perpendicular motions~\cite{Nazarov2013,Wicki-16}, and it
behaves qualitatively differently from the 1D bands: there are no $T$-resonances,
and the transit time is approximately proportional to the thickness $d$. In the
band gaps transmission is effected by evanescent waves, so both the dwell time
$\TDW$ and self-interference time $\TSI$ [Eq.~(\ref{eq:IDWINFUL})] saturate
with $d$, and so does the transit time $\TTT=\DTW+\TFM$: in the gap center
$\TTT^\infty=180$~as in the lower and 190~as in the upper gap.

\begin{figure}[b] %%%%%%%%%%%%%%%%%%%%%%%%%%%%%%%%%%%%%%%%%%%%%%%%%%%%%%%%%% F4
\centering
\includegraphics[trim = 0mm 0mm 00mm 0mmm,width=0.48\textwidth]{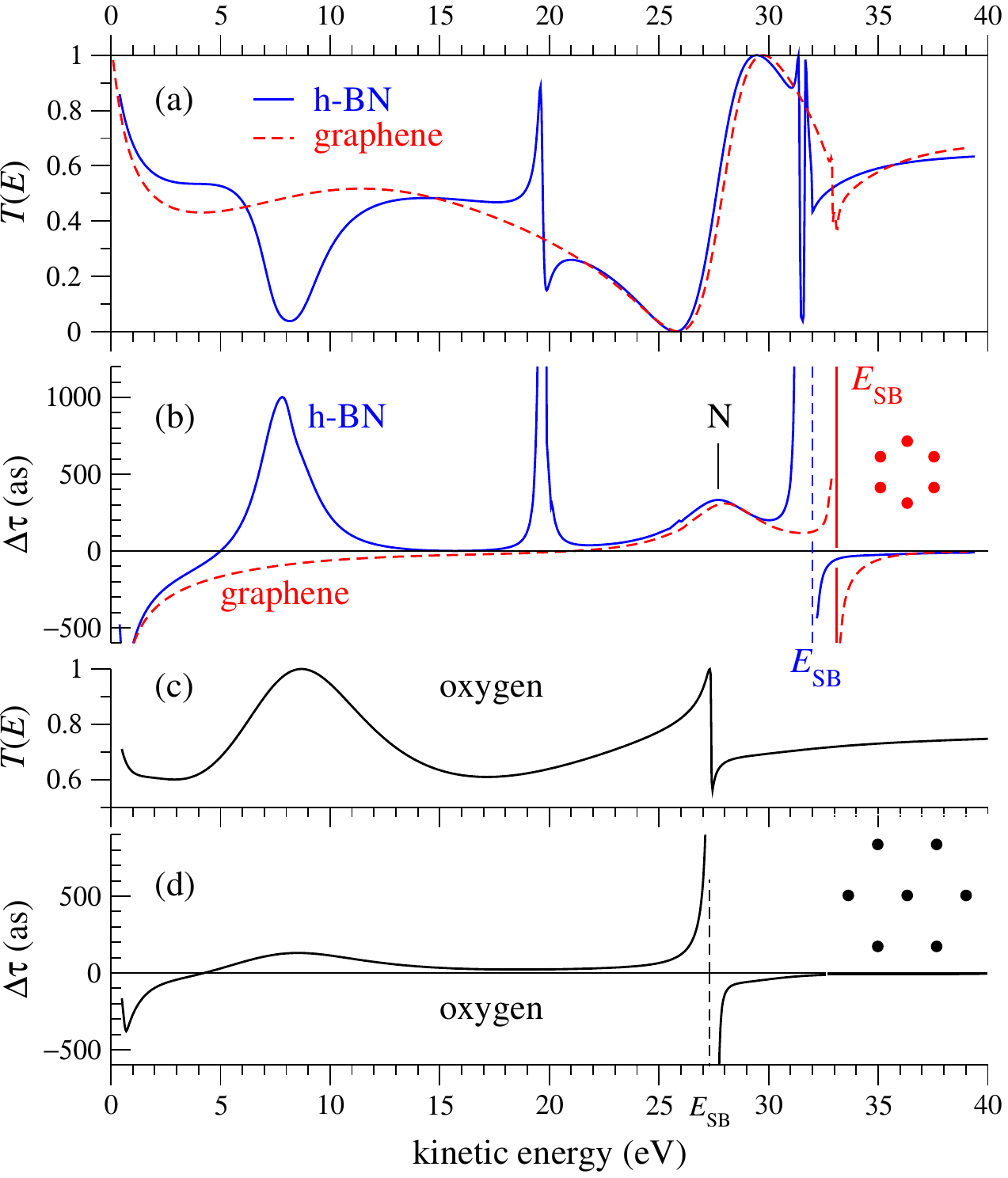} 
\caption{(a)~Transmissivity $T(E)$ of 1ML graphene [red dashed line, same as the upper
curve in Fig.~\ref{f:CBS}(a)] compared to 1ML h-BN (blue). (b)~Wigner delay $\DTW$ for graphene
(red) and h-BN (blue). (c)~$T(E)$ for 1ML oxygen (hexagonal structure~\cite{Krasovskii_2015}).
(d)~$\DTW$ for 1ML oxygen. Vertical bars indicate critical energies $\ESB$. Circles in
the insets show fragments of the lattices of graphene (red) and oxygen (black).
}
\label{f:neg}
\end{figure}

We have seen in Fig.~\ref{f:transit} that the dwell time $\TDD$ diverges on approaching
$\ESB$ leading to a divergence of the Wigner time delay. Let us now focus on energies just
above $\ESB$. There the simple relation $\XT_0-\XR_0=\pm\pi/2$ does not hold, and the general
formula~(\ref{eq:GENID3D}) should be used. We can write $\XT_0-\XR_0=\gamma(E)\pm\pi/2$, with
$\gamma(\ESB)=0$, and it follows from Eq.~(\ref{eq:IDCOS}) that 

\begin{equation}
k_0|t_0r_0|\sin(\gamma)=
\sum\limits_{0<|\Gv|< k_0} k_\Gv|t_\Gv r_{\hat s\Gv}|\cos(\XT_\Gv-\XR_{\hat s\Gv}).
\label{eq:IDCOS1}
\end{equation}
In the vicinity of $\ESB$, owing to the steeply growing $k_\Gv$, the energy derivative of
$\gamma$ diverges as $1/\sqrt{(\ESB-E)E}$. However, unlike the case of $E<\ESB$, the sign of
the divergent term may be different in different systems. This divergence turns out to lead
to negative time delay in particular in graphene, h-BN, and a monolayer oxygen in the geometry
of ruthenium surface oxide~\cite{Krasovskii_2015}, see Fig.~\ref{f:neg}.

Above the respective $\ESB$ energies the three monolayers have similar shape of both $T(E)$
and $\DTW(E)$ curves, with $\DTW(E)$ diverging toward $-\infty$.
This nontrivial speed-up cannot be predicted from the general
theory, so we have validated its consistency by comparing $\DTW$ obtained as
$\dot\eta_0$ with the value derived from the phase derivatives of the propagating
secondary beams $\dot{\XR_\Gv}$ and $\dot{\XT_\Gv}$ and the density integral
$\TDD$ through Eq.~(\ref{eq:GENID3D}). For all three systems the discrepancies
in $\DTW$ are negligible over the entire range considered, see Fig.~\ref{f:acc}
%S2 in SM~\cite{Suppl}.

In the language of the Hartman effect, i.e., if one assumes that the wave packet can be
ascribed a trajectory, unlimited negative time delay implies negative transit time, which
illustrates that $\TTT$ has no physical meaning. Nevertheless, for a sufficiently spectrally
narrow wave packet this effect can be measured as a substantial reduction of the arrival time,
the essential novel aspect being that no stop band is involved, and consequently the
transmitted fraction of the current is orders of magnitude larger than in tunneling---above
50\%, see Figs.~\ref{f:neg}(a) and~\ref{f:neg}(c). A spectrally wide packet with energy around
$\ESB$ will be torn into two spatially separated ones with comparable intensities.

The $\DTW$-divergence due to the emergence of the secondary beams is a general
  phenomenon, with $\ESB$ depending solely on the surface Bravais lattice, so it is to be
  expected for any exfoliated material. Apart from that, the atomic structure of a specific
  multilayer may bring about additional strong features: One example is the broad $\DTW$
  maximum at 28~eV due to the N-resonance in graphene and h-BN. Furthermore, the more
  complicated geometry of h-BN gives rise to a deep $T(E)$ minimum at 8~eV and a sharp
  resonance at 20~eV [manifested by a steep drop of $T(E)$], see Fig.~\ref{f:neg}(a), both
  structures manifesting a significant increase of $\DTW$, see Fig.~\ref{f:neg}(b). The absence
  of the N-resonance in oxygen monolayer, Fig.~\ref{f:neg}(c), is explained by its simple
  hexagonal geometry in contrast to the honeycomb lattice of graphene and h-BN. However, the
  oxygen monolayer manifests the same type of divergence at $\ESB$ as the other materials. One
  may expect the more complicated 2D structures, such as transition metal dichalcogenids,
  to show more interesting features.

The general formula~(\ref{eq:IDCOS1}) proves the divergence of the derivative
of the phase difference between transmission and reflection. We will now present
a simple 2D model, for which the divergence of the phase derivatives $\dot\XT_0$
and $\dot\XR_0$ themselves can be demonstrated analytically. Consider an infinite
chain of atoms along $\hat{\mathbf x}$ axis modeled by a $\delta$-function
potential $\Omega\delta(y)$ in the $\hat{\mathbf y}$ direction and a weak
corrugation $\cos(\GSB x)$ along $\hat{\mathbf x}$:
$\hat{H}=(\hat p_x^2+\hat p_y^2)/2m+2(\hbar^{2}/m)\Omega\delta(y)\cos(\GSB x)$.
The $\cos(\GSB x)$ perturbation couples the motions along $\hat{\mathbf x}$ and
$\hat{\mathbf y}$ and gives rise to the secondary beams at $\ESB = \hbar^{2}\GSB^2/2m$.

By applying the Lippmann-Schwinger equation to the Laue representation of the wave function
for normal incidence (along $\hat{\mathbf y}$), $\Psi(x,y) = \sum_{g}\phi_{g}(y)\exp(igx)$,  
we obtain the central beam wave function in terms of the Green's function for a 1D free
motion $\GFA(y;\omega)$ \cite{Economou}:
\begin{equation}
\phi_{0}(y)=\exp({iky})-\frac{2i\Omega\exp({ik|y|})}{k(F_{\GSB }^{-1}-2F_{0}-F_{2\GSB })},
\label{eq:phi0y}  
\end{equation}
where $F_{g} \equiv \hbar^2\Omega\GFA(0;k^2-g^2)/m$, see Appendix \ref{s:model}. 
 %~\cite{Suppl}.
Equation~(\ref{eq:phi0y}) yields the reflection and transmission amplitudes:
$r=-2i\Omega /[k(F_{\GSB }^{-1}-2F_{0}-F_{2\GSB })]$ and $t=1+r$. Around $\ESB$, it
can be proved that $F_{\GSB }^{-1}=-\sqrt{\GSB^2-k^2}/\Omega$
for $\GSB^2>k^2$ and $F_{\GSB }^{-1}=i\sqrt{k^2-\GSB^2}/\Omega$ for $\GSB^2<k^2$,
see Appendix \ref{s:model}. %Sec.~III of SM~\cite{Suppl}.
The other two parameters $F_g$ in Eq.~(\ref{eq:phi0y}) become
$F_{0}=-i\Omega/\GSB $ and $F_{2\GSB }=-\Omega/(\GSB \sqrt{3})$, so $\DTW$ becomes
\begin{align}
\DTW \approx \frac{m}{\hbar \beta}\times \begin{cases}
\dfrac{6\GSB^2\Omega^2}{13\left(\Omega^2-\beta\sqrt{3}\right)^2}, & E<\ESB,
\\[15pt]
\dfrac{12\sqrt{3}\GSB^2\Omega^4\left(\Omega^2+\beta\right)}
{13\left[\Omega^4+3\beta
\left(\Omega^2+\beta\right)\right]^2}, & E>\ESB,
\end{cases}\label{eq:tauW2D}
\end{align}
where $\beta\equiv(2m/\hbar^2)\sqrt{\ESB|E-\ESB|}$
yields the divergence of the Wigner time delay, cf. Eq.~(\ref{eq:IDCOS1}).

To summarize, our study has revealed hitherto unknown phenomena caused by the
in-plane umklapp scattering, which are especially conspicuous in monolayers:
the steep increase of the delay at the transmission minima in h-BN and, most
strikingly, the divergent delay followed by divergent advance of the central
beam due to the emergence of the secondary beams observed in the monolayers
of graphene, h-BN, and oxygen. In thicker multilayers, the lateral scattering
damps the sharp $T$-resonances and eliminates the related huge time delays.

The rapid energy variations of the time delay cause a spatial reshaping of the
wave packet, which can be employed to experimentally study the details
of the acceleration of the outgoing photoelectrons by a spatially inhomogeneous
infrared laser field at the surface. This is an insufficiently studied question,
both experimentally and theoretically, and the usual assumption is to neglect the
spatial extent of the outgoing wave packet~\cite{Cavalieri2007,
Schultze10,Neppl2012,Neppl15,Okell:15,Siek2017,Ossiander2018,Riemensberger2019}.
The possibility to split the wave packet into two---separated by a fraction of a
femtosecond---and observe their acceleration by the same probe pulse opens a way
to clarify the details of the streaking mechanism, which is instrumental for the
interpretation of the measurements. Such a possibility is offered by a thin overlayer
in which the time delay rapidly changes with energy due to a specific crystal
structure (as in h-BN) or due to the emergence of the secondary beams, common to all
crystalline films. The advantage over the traditional Hartman effect is that the
in-plane scattering induced $\DTW$ resonances are not accompanied by a dramatic
drop of the transmitted flux. In addition, owing to the minimal thickness of a
monoatomic overlayer, the streaking field is minimally disturbed. One may argue
that the phase time that underlies the present study is an asymptotic value, and
the extrapolation of the $z\to\infty$ equation of motion to the vicinity of the
scatterer is not rigorously justified~\cite{Hauge1989,Hauge1997} (even though this
is quite common, see, e.g., Refs.~\cite{Eckle2008,Sainadh2019,Camus2017} and an
analysis in Ref.~\cite{Torlina2015}). Still, our numerical streaking experiment~\cite{Kuzian20}
showed that the photoelectron escape time extracted from streaking reflected all
the band-structure related features of the relevant phase time, and even the
absolute values were reasonably close. Thus, our present results offer a promising
starting point to study the wave-packet speed-up unrelated to tunneling.

Apart from the practical applications, the discovered properties of the ultrathin layers
may be interesting from the point of view of developing quantum-mechanical formalisms in
which the motion of the quantum particle is considered in terms of
trajectories~\cite{Oriols1996,Gruebl2002,Leavens2008}.

%\vfill

\begin{acknowledgments}
This work was supported by the Spanish Ministry of Science and Innovation
(MICINN Project No.~PID2022-139230NB-I00) and by the National Academy
of Sciences of Ukraine (Project No.~III-2-22 and III-4-23). 
\end{acknowledgments}

\appendix
\begin{figure}[htb] %%%%%%%%%%%%%%%%%%%%%%%%%%%%%%%%%%%%%%%%%%%%%%%%%%%%%%%%%% F2 NC6
\centering
\includegraphics[trim = 0mm 0mm 00mm 0mmm,width=0.48\textwidth]{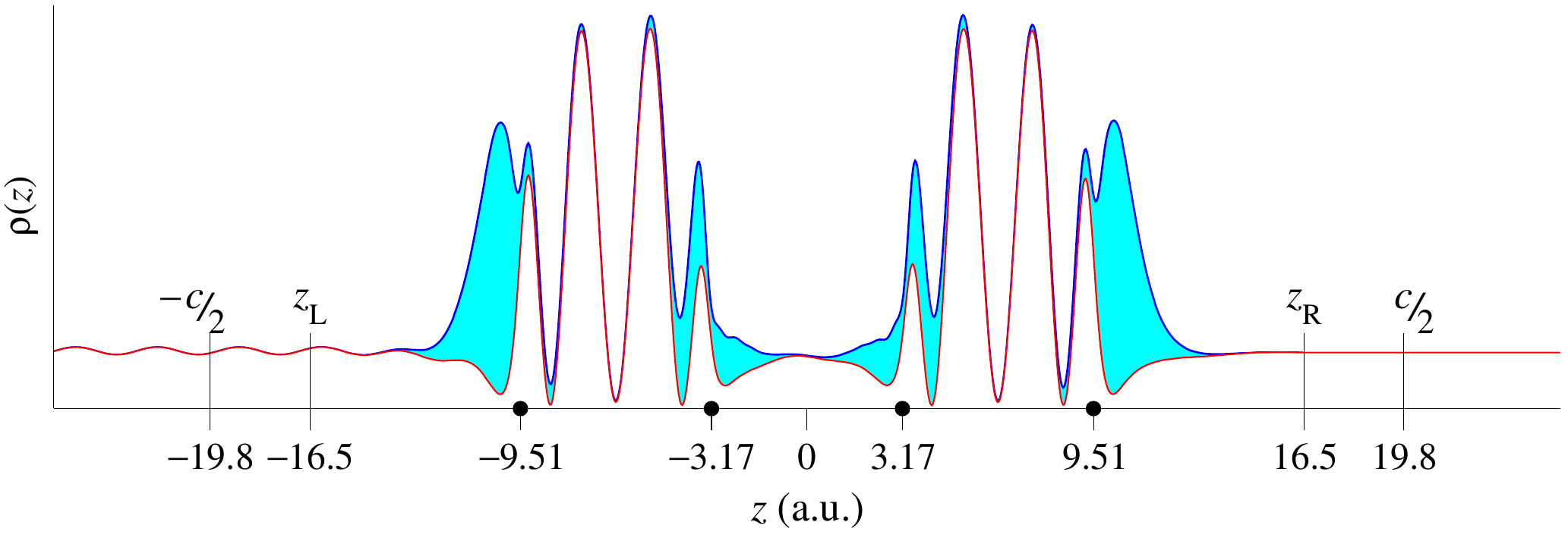} 
\caption{Probability density profile $\rho(z)$ of the scattering wave function for
  normal incidence on a 4ML graphene film at 18.24~eV kinetic energy, corresponding
  to a transmission resonance, see Fig.~\ref{f:acc}. Blue line is the total density and
  red line is the $\Gv=0$ contribution $|\phi_0(z)|^2$, see Eq.~(\ref{eq:Laue}). The cyan
  shaded area shows the contribution from the $\Gv\ne0$ Fourier harmonics. The
  artificial supercell extends from $-c/2$ to $c/2$. Black circles show the location
of the four graphene planes.}
\label{f:scheme}
\end{figure}
\section{Calculation of scattering states}\label{s:method}
The crystal potential
is constructed within the local density approximation; it includes both the Coulomb
singularities at the nuclei and a realistic full-potential shape everywhere between
$\ZL$ and $\ZR$. In the scattering region an all-electron wave function $\Psi$
is a linear combination of the Bloch eigenfunctions of an auxiliary band structure
problem for a periodic supercell extending from $-c/2$ to $c/2$, see Fig.~\ref{f:scheme}. Then
the wave function is Fourier expanded to arrive at the Laue representation %%%~(\ref{eq:Laue}),
\begin{equation}
\Psi(\rV,z) = \sum\limits_{\Gv} \phi_{\Gv}(z)\exp[\,i\Gv\rV\,],
\label{eq:Laue}
\end{equation}
%
%%% \noindent
so the density profile $\rho(z)$ of the scattering state is
$\rho(z)=\sum\limits_{\Gv}|\phi_{\Gv}(z)|^2$.
%%* \begin{equation}
%%* \rho(z) = \int|\,\Psi(\rV,z)\,|^2\, d\rV = \sum\limits_{\Gv}|\phi_{\Gv}(z)|^2.
%%* \label{profile}
%%* \end{equation}
%%
An example for the normal incidence on a 4ML graphene slab at $E=18.24$~eV
is shown in Fig.~\ref{f:scheme}. Some more details of the application of the method to
graphene monolayer and bilayer have been presented in
refs.~\cite{Nazarov2013,Krasovskii21,Krasovskii22}.
\begin{figure}[htb] %%%%%%%%%%%%%%%%%%%%%%%%%%%%%%%%%%%%%%%%%%%%%%%%%%%%%%%%%% F3
\centering
\includegraphics[trim = 0mm 0mm 00mm 0mmm,width=0.48\textwidth]{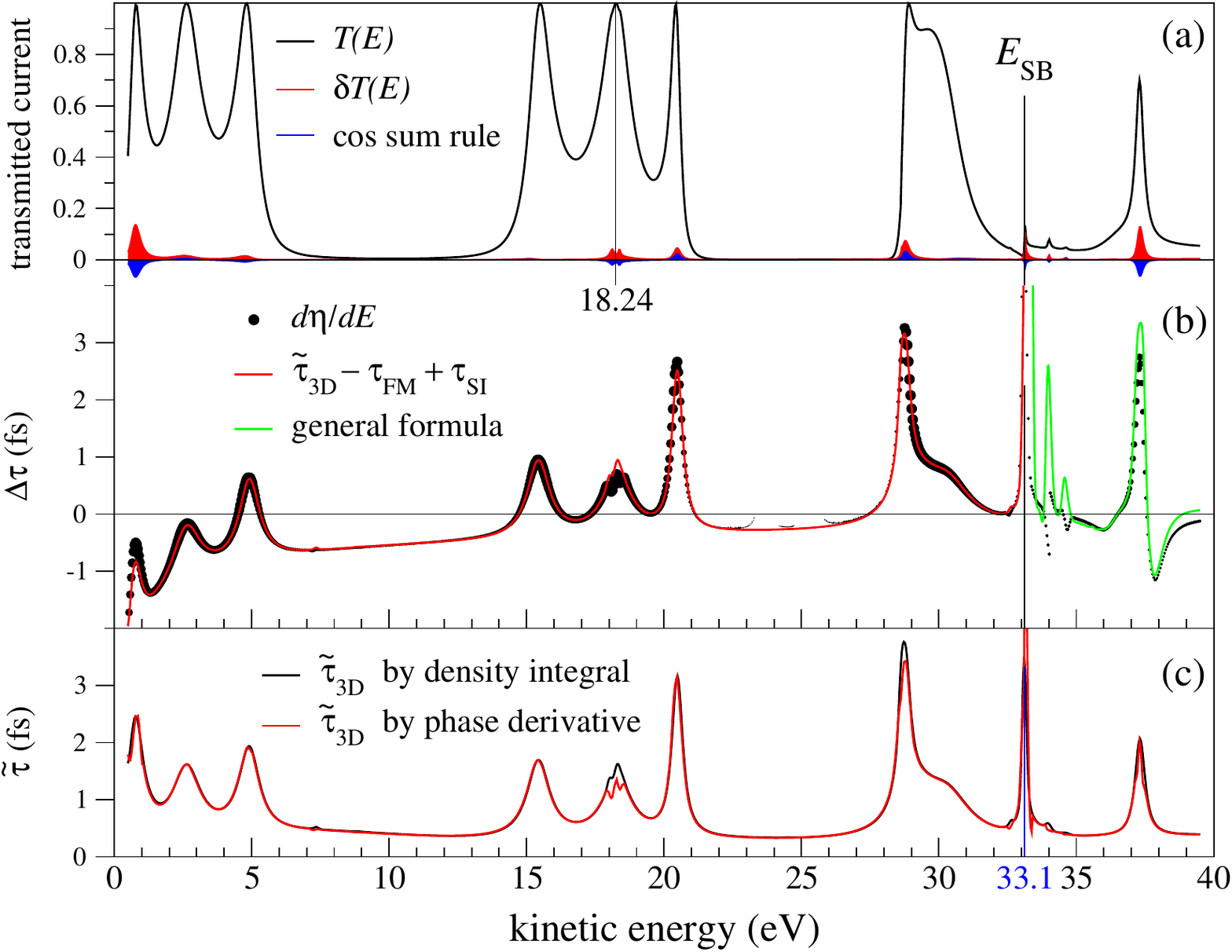} 
\caption{(a)~Total current $T(E)$ through 4ML graphene (black curve), current 
  nonconservation $\delta T$ (red shading), and sum rule~(3) %%%% (\ref{eq:IDCOS})
  (blue shading). (b)~Delay $\DTW$ as the energy derivative $\dot\eta$ (black circles),
  from the analogue of Eq.~(5), $\TDD=\DTW+\TFM-\TSI$, below $\ESB$ (red), and from the
  general formula~(7) above $\ESB$ (green). Size of the circle is proportional
  %%% (\ref{eq:GENID3D})
  to $|t_0|$. Below $\ESB$ agreement between the two curves depends on the phase
  relation $\XT_0-\XR_0=\pm\pi/2$. (c)~Dwell time~(6) %%% (\ref{eq:DEFDWELL3D})
  (black) compared to the right hand side of Eq.~(7) (red). %%% (\ref{eq:GENID3D})
}
\label{f:acc}
\end{figure}

The variational method for scattering~\cite{Krasovskii2004} is based on minimizing the
functional $\|(\hat H -E)\Psi\|$, so the solution~(\ref{eq:Laue}) satisfies the Schr\"odinger
equation only with certain accuracy. Identities~(2)--(7) allow
%%%% ~(\ref{eq:IDCF})--(\ref{eq:GENID3D})
us to estimate the computational uncertainty and verify that it is much smaller than the
physically relevant quantities. Figure~\ref{f:acc}(a) shows the total current spectrum

\begin{equation}
T(E)=\sum\limits_{|\Gv|<\KKO}T_\Gv(E)\KKG/\KKO
\label{eq:T}
\end{equation}
%
%%% \noindent
together with the current nonconservation $\delta T(E)$ defined as the difference  of
the probability fluxes at $\ZL$ and $\ZR$. The parameter $\delta T$
is a good indicator of the overall quality of the wave function: for 4MLs its average
value over the range of 40~eV is below 1\%, and it sharply peaks to exceed 5\% in
a few narrow intervals, accompanied by a slight violation of the identities~(3), (6), or (7).

%%%(\ref{eq:IDCOS}), (\ref{eq:DEFDWELL3D}), or (\ref{eq:GENID3D}).
Computationally, dwell time $\TDW$ is the most reliable quantity, so below $\ESB$
the delay $\DTW$ can be obtained without resorting to numerical differentiation. This
is especially important when $T_0$ drops below~$10^{-6}$, whereby its phase $\XT_0$ becomes
highly unreliable, as, e.g., between 22 and 27~eV in Fig.~\ref{f:acc}(b). The uncertainties
are of similar magnitude for the 3, 5, and 6MLs, and they are much smaller (practically
negligible) for 1 and 2MLs. Thus, the accuracy of $\DTW$ is sufficiently high to
enable a detailed analysis of scattering by multilayers.

\section{Phases and dwell time in 3D case}\label{app:ID}
According to Eq.~(1) of the main text, the wave function for a wave
incident %(\ref{eq:tr})
from the left in the left half-space reads
\begin{flalign}\nonumber
\Psi_{\rightarrow}(\rV,z) =\exp{{ik_0z}}+\sum\limits_\Gv r_\Gv\, \exp{i[-k_\Gv(z-\ZL)+\Gv\rV]}. &&
\end{flalign}
%
%~(\ref{eq:tr}) 
Also, it follows form Eq.~(1) that for a symmetric crystal the wave incident from
the right in the left half-space is 
\begin{flalign}\nonumber
\Psi_{\leftarrow}(\rV,z) = \sum\limits_\Gv t_\Gv\, \exp{i[-k_\Gv(z-\ZL)+\Gv\rV]}.&&
\end{flalign}
The superposition $\PSL=\Psi_{\rightarrow}+\Psi_{\leftarrow}$ carries zero current, and in vacuum
the condition ${\rm Re}\int\!\PSL^*(-i\partial\PSL/\partial z)\,d\rV=0$ results in the relation
\begin{flalign}\label{eq:appCF}
k_0[r_0t_0^*+t_0r_0^*]+\sum\limits_{0<|\Gv|<k_0}k_\Gv|t_\Gv+r_\Gv|^2&=0, 
\end{flalign}
where the sum is over the propagating secondary beams. This leads to Eq.~(5).
%(\ref{eq:IDCF}).
Below $\ESB$ equation~(\ref{eq:appCF}) reduces to ${\rm Re}\,(t_0r_0^*)=0$, which implies
$\cos(\XT_0-\XR_0)=0$ or $\XT_0-\XR_0=\pm\pi/2$, as in the 1D case~\cite{Falck1988}.
For several propagating beams Eq.~(\ref{eq:appCF}) leads to Eq.~(6). %(\ref{eq:IDCOS}).

Next we derive the expression for the dwell time in the 3D case. We follow the
original derivation by Smith~\cite{Smith60}: From the Schr\"odinger equation
$\hat H\Psi=E\Psi$ and its energy derivative $(\hat H-E)\dot\Psi=\Psi$ we obtain
\begin{equation}\label{eq:appLAPLACE}
-\frac{\hbar^2}{2m}(\Psi^*\Delta\dot\Psi-\dot\Psi\Delta\Psi^*)=\Psi^*\Psi.
\end{equation}
After integrating over $\rV$, the $\rV$ part of the Laplacian in the left-hand
side vanishes,
$\sum\limits_\Gv(\phi_\Gv^*\dot\phi_\Gv-\dot\phi_\Gv\phi_\Gv^*)|\Gv|^2=0$, and in
the Laue representation~(\ref{eq:Laue}) equation~(\ref{eq:appLAPLACE}) becomes
\begin{equation}\label{eq:Smith18}
-\frac{\hbar^2}{2m}\sum\limits_\Gv\frac{d}{dz}(\phi_\Gv^*\dot\phi_\Gv'-\dot\phi_\Gv\phi_\Gv'^*)
=\rho(z),
\end{equation}
where the prime stands for the derivative $d/dz$ and $\rho(z)$ is the density distribution
profile. Integrating Eq.~(\ref{eq:Smith18}) from $\ZL$ to $\ZR$ we obtain
\begin{equation}\label{eq:Smith19}
  -\frac{\hbar^2}{2m}\left.\sum\limits_\Gv(\phi_\Gv^*\dot\phi_\Gv'
  -\dot\phi_\Gv\phi_\Gv'^*)\;\right|_{\ZL}^{\ZR}=
  \int\limits_{\ZL}^{\ZR}\!\!\rho(z)\,dz\equiv Q.
\end{equation}

Let us express the boundary values in terms of the notation of Eqs.~(1) %(\ref{eq:tr})
and~(2) %(\ref{eq:XITR}):
\begin{widetext}
\begin{equation}
  \begin{array}{l}
    \begin{aligned}
  \text{left boundary: }\phi_\Gv^*(\ZL)&=\delta_{0\Gv}+\RRG^*, \\
\dot\phi_\Gv(\ZL)& =\RDG, \\
\dot\phi_\Gv'(\ZL)&=i\left(\delta_{0\Gv}\KDO-\KKG\RDG-\RRG\KDG\right), \\ 
\phi_\Gv'^*(\ZL)&=i\left(-\delta_{0\Gv}\KKO+\RRG^*\KKG^*\right),
\end{aligned}
\end{array}
  \begin{array}{l}
    \begin{aligned}
    \text{right boundary: }\phi_\Gv^*(\ZR)&=\TTG^*, \\
\dot\phi_\Gv(\ZR) &=\TDG, \\
\dot\phi_\Gv'(\ZR)&=i\left(\KKG\TDG+\TTG\KDG\right), \\ 
\phi_\Gv'^*(\ZR)&=-i\TTG^*\KKG^*,
\end{aligned}
\end{array}
\end{equation}
where $\KKG^*=\KKG$ for propagating beams and $\KKG^*=-\KKG$ for evanescent beams.
Then Eq.~(\ref{eq:Smith19}) becomes
\begin{equation}\label{eq:LONG}
\frac{2m}{\hbar^2}Q=i\KDO +2\KDO{\rm Im}\,\RRO -
  i\!\!\sum\limits_{|\Gv|<\KKO}2(\RDG\RRG^*+\TDG\TTG^*)\KKG + (R_{\Gv}+T_{\Gv})\KDG -
  i\!\!\sum\limits_{|\Gv|>\KKO}(R_{\Gv}+T_{\Gv})\KDG, 
\end{equation}
where $R_{\Gv}=\RRG\RRG^*$ and $T_{\Gv}=\TTG\TTG^*$ are partial reflection and transmission
probabilities. From the current conservation law
$\dfrac{d}{dE}\sum\limits_{|\Gv|<\KKO}(\RRG\RRG^*+\TTG\TTG^*)\KKG=\KDO$
it follows that $\sum\limits_{|\Gv|<\KKO}(\RDG\RRG^*+\TDG\TTG^*)\KKG+(R_{\Gv}+T_{\Gv})\KDG = 
\KDO-\sum\limits_{|\Gv|<\KKO}(\RRG\RDG^*+\TTG\TDG^*)\KKG$,
%
%%% \begin{flalign}\label{eq:CCL1}
%%% \sum\limits_{|\Gv|<\KKO}(\RDG\RRG^*+\TDG\TTG^*)\KKG+(R_{\Gv}+T_{\Gv})\KDG = 
%%% \KDO-\sum\limits_{|\Gv|<\KKO}(\RRG\RDG^*+\TTG\TDG^*)\KKG,
%%% \end{flalign}
%
which we substitute into the $|\Gv|<\KKO$ sum of (\ref{eq:LONG}) to obtain
\begin{align}\label{eq:LONG1}
  \frac{2m}{\hbar^2}Q= 2\KDO{\rm Im}\,\RRO -
  i\!\!\sum\limits_{|\Gv|<\KKO}(\RDG\RRG^*+\TDG\TTG^*-\RDG^*\RRG-\TDG^*\TTG)\KKG
 -i\!\!\sum\limits_{|\Gv|>\KKO}(R_{\Gv}+T_{\Gv})\KDG.
\end{align}
Using the identity
${\rm Im}\,(\RDG\RRG^*+\TDG\TTG^*)=R_{\Gv}\dot{\XR_{\Gv}}+ T_{\Gv}\dot{\XT_{\Gv}}$
and the derivatives of the wave vectors $\KDG=\dfrac{m}{\hbar^2\KKG}$, we obtain 
\begin{align}\label{eq:LONG4}
 \TDW= \frac{{\rm Im}\,\RRO}{\KKO\VVO} +
 \frac{\hbar}{\VVO} \sum\limits_{|\Gv|<\KKO}(R_{\Gv}\dot{\XR_{\Gv}}+T_{\Gv}\dot{\XT_{\Gv}})\VVG
 -\frac{1}{\VVO} \sum\limits_{|\Gv|>\KKO}\frac{R_{\Gv}+T_{\Gv}}{2|\KKG|},
\end{align}
\end{widetext}
where for the evanescent waves $|\KKG|=\sqrt{\hbar^2|\Gv|^2/2m-E}$.
The $|\Gv|>\KKO$ sum in Eq.~(\ref{eq:LONG4}) is the integral probability density
stored in the evanescent tails of $\Psi$ in both half-spaces outside the region
technically assigned to the scatterer. Thus, it is natural to unite the last term in
Eq.~(\ref{eq:LONG4}) with $\TDW$ and introduce $\TDD$, see Eq.~(6), %(\ref{eq:DEFDWELL3D}),
which yields Eq.~(7) %(\ref{eq:GENID3D}).

%%%* \subsection{2D Nearly free electron model analysis}
%%%*  It gives Eq. (\ref{eq:Fi0xy}).

\begin{widetext}
\section{2D Nearly-free-electron model analysis}\label{s:model}
We consider the Hamiltonian %%%*.
\begin{equation}
\hat{H}=\frac{\hat{p}_{x}^{2}+\hat{p}_{y}^{2}}{2m}
+2\frac{\hbar^{2}}{m}\Omega\delta(y)\cos(\GSB x). \label{eq:H2Da}
\end{equation}
%
%%% \noindent
The Lippmann--Schwinger equation for the wave function for a normal incidence
(along $\hat{\mathbf y}$) on the chain reads 

\begin{flalign}
& \Psi(x,y) = \exp{iky}\; +\frac{\hbar^{2}}{m}\int\!\! dx'\GFB
\left(x-x',y;k^2\right)2\Omega\cos(\GSB x')\Psi(x',0),  && \label{eq:LipSha} %%% && \nonumber
\end{flalign}
\begin{flalign}
&\text{where}\;\; \GFB\left(x,y;\omega\right)
=\frac{2m}{\hbar^{2}}\iint\frac{dq_{x}dq_{y}}{4\pi^{2}}\frac{\exp(iq_{x}x+iq_{y}y)}
{\omega-q_x^2-q_y^2} && \nonumber
\end{flalign}
%
%%% \noindent
is the Green's function for a 2D free motion. Substituting the Laue
representation (\ref{eq:Laue})
in both sides of Eq.~(\ref{eq:LipSha}) yields
%
%\begin{widetext}
\begin{flalign}
 \Psi(x,y) =& \exp(iky)\, && \nonumber \\ 
 +& \frac{\hbar^{2}}{m}\Omega\sum_{g}\phi_{g}\left\{\exp[i(g+\GSB)x]\GFA
\left[y;k^{2}-\left(g+\GSB \right)^{2}\right] %%% \right.\nonumber && \\ & \left.
+\exp[i\left(g-\GSB \right)x]\GFA\left[y;k^{2}-(g-\GSB )^{2}\right]\right\},&&
\label{eq:Laue2D}
\end{flalign}
%
%%% \noindent
where $\phi_{g}\equiv\phi_{g}(y=0)$ and $\GFA(y;\omega)$ is the Green's
function for a free motion in one dimension~\cite{Economou},

\begin{align}
\GFA(y,\omega) & =\begin{cases}
-\dfrac{m}{\hbar^{2}q}\exp{ -q |y|}, & \omega<0,\\[12pt]
-\dfrac{im}{\hbar^{2}q}\exp{iq |y|}, & \omega>0, 
\end{cases}\quad\label{eq:g01D}  
\end{align}
%
%%% \noindent
where $q \equiv\sqrt{|\omega|}$.
It follows from Eq.~(\ref{eq:g01D}) that for $y\rightarrow \pm\infty$, in the Laue
representation~(\ref{eq:Laue2D}) of $\Psi(x,y)$ one can retain only propagating terms,
i.e., those with the positive $\omega$ argument of $\GFA$. Then, around $\ESB$, we have 
\begin{flalign}
&\Psi(x,y)  =
  e^{iky}+\frac{\hbar^{2}}{m}\Omega
  \left\{(\phi_{\GSB} +\phi_{-\GSB })\GFA(y;k^2) +[2\phi_{0}\cos(\GSB x)
      +\phi_{2\GSB }e^{i\GSB x}+\phi_{-2\GSB }e^{-i\GSB x}]\GFA(y;\epsilon)\right\}, &&
  \nonumber
\end{flalign}
%
%%% \noindent
\end{widetext}
where $\epsilon\equiv k^2-\GSB^2$. The central beam scattering is given by
the first term in the rectangular brackets.
\begin{equation}
\phi_{0}(y)=\exp({iky})-\frac{i\Omega\exp({ik|y|})}{k}(\phi_{\GSB }+\phi_{-\GSB }).
\label{eq:Fi0xy}
\end{equation}
%
%%% \noindent
Equating the coefficients of the Fourier harmonics $\exp(igx)$ in Eq.~(\ref{eq:Laue2D})
we obtain

\begin{align}
\phi_{g} & =\delta_{g0}+\left(\phi_{g-\GSB }+\phi_{g+\GSB }\right)F_{g},
\label{eq:recfig}\\
F_{g} & \equiv \frac{\hbar^{2}}{m}\Omega\GFA\left(0;k^2-g^2\right). \label{eq:fg}
\end{align}
%
%%% \noindent
If the potential is small compared with the kinetic energy $\ESB$ and oscillates
sufficiently rapidly, $\Omega/\GSB\ll 1$, a nearly-free-electron approximation may
be applied, i.e., only the shortest-$g$ terms retained in the Laue representation:
$g=0,\pm\GSB,\pm2\GSB$. 
By truncating the chain of equations (\ref{eq:recfig}) at $\phi_{\pm2\GSB }$ we find 

\begin{equation}
\phi_{\GSB }=\phi_{-\GSB }\approx\frac{1}{F_{\GSB }^{-1}-\left(2F_{0}+F_{2\GSB }\right)}.
\label{eq:fiK}
\end{equation}

Finally, we substitute $\phi_{\pm\GSB }$ into Eq. (\ref{eq:Fi0xy}) to obtain
\begin{equation}
\phi_{0}(y)=\exp({iky})-\frac{2i\Omega\exp({ik|y|})}{k(F_{\GSB }^{-1}-2F_{0}-F_{2\GSB })},
\label{eq:phi0y}  
\end{equation}
which is Eq.~(9) of the main text. Equation~(\ref{eq:phi0y})
yields the reflection and transmission amplitudes:
$r=-2i\Omega /[k(F_{\GSB }^{-1}-2F_{0}-F_{2\GSB })]$ and $t=1+r$.
In the vicinity of $\ESB$ it holds $\epsilon/\GSB^2 \ll 1$, and we may set
$\epsilon=0$ in all terms except
\begin{flalign}
F_{\GSB }^{-1}= \begin{cases}  -\sqrt{|\epsilon|}/\Omega, &\epsilon<0, \\
                             \quad\, i\sqrt{\epsilon}/\Omega, &\epsilon>0.
\end{cases}
\end{flalign}
Then, in the limit $k\to\GSB$ the coefficients $F_g$ in Eq.~(\ref{eq:phi0y}) become
$F_{0}\to-i\Omega/\GSB $ and $F_{2\GSB }\to-\Omega/(\GSB \sqrt{3})$, and 
we obtain expressions~(10) for $\DTW$, which demonstrate the divergence
of the Wigner time delay for $\epsilon \to 0$, i.e. for $E \to\ESB$. 

\bibliography{Hartman}

\end{document}